\documentclass[]{llncs}
\usepackage[utf8]{inputenc}
\usepackage{tikz}
\usepackage{amsmath}
\usepackage{mathtools}
\usepackage[hyphens]{url}
\usepackage[colorlinks=true,urlcolor=black]{hyperref}

\title{Collaborative Deanonymization}
\author{Patrik Keller\inst{1} \and Martin Florian\inst{2} \and Rainer Böhme\inst{1}}
\institute{%
University of Innsbruck\\
\email{\{patrik.keller, rainer.boehme\}@uibk.ac.at}\and
Weizenbaum Institute, Humboldt University of Berlin\\
\email{martin.florian@hu-berlin.de}
}
\begin{document}
\maketitle
\providecommand{\tx}{\ensuremath{T}} %
\providecommand{\mixin}{\ensuremath{m}} %
\providecommand{\perm}{\ensuremath{{\psi}}} %
\providecommand{\select}{\ensuremath{{\sigma}}} %
\providecommand{\target}{\ensuremath{{t}}} %
\providecommand{\LEA}{LEA}
\providecommand{\cj}{CoinJoin}

\DeclarePairedDelimiter\set{\{}{\}}
\DeclarePairedDelimiter\len{\lvert}{\rvert}

\newcommand{\Aq}{\vphantom{Aq}}

\newcommand{\todo}[1]{\textcolor{gray}{\emph{#1}}} %

\begin{abstract}
  Privacy-seeking cryptocurrency users rely on anonymization techniques like \cj{} and ring transactions.
  By using such technologies benign users potentially provide anonymity to bad actors.
  We propose overlay protocols to resolve the tension between anonymity and accountability in a peer-to-peer manner.
  Cryptocurrencies can adopt this approach to enable prosecution of publicly recognized crimes.
  We illustrate how the protocols could apply to Monero rings and \cj{} transactions in Bitcoin.
\end{abstract}
\section{Introduction}

``Anonymity loves company.''~\cite{dingledine2006Anonymityloves}
It is well-established that anonymity is co-created by the members of an anonymity set, who share the same intention and employ technical systems and protocols to make them appear indistinguishable to outside observers~\cite{pfitzmann2001AnonymityUnobservability}.
Inherently, benign members seeking privacy assist bad actors avoiding law enforcement.

Previously, the tension between privacy and law enforcement has been studied for mixes in communication networks~\cite{claessens2003Revocableanonymous,kopsell2006RevocableAnonymity,backes2014BackRefAccountability}.
The proposed solutions rely on putting backdoors into systems or the supporting cryptography, such that designated parties can revoke the anonymity in justified cases.
Access to the backdoor is made transparent, which holds law enforcement accountable and impedes mass surveillance.
With the advent of privacy-hardened cryptocurrencies, the tension is instantiated for money flows.
While backdoors seem technically feasible, it is unlikely that they can be sustained in decentralized systems, whose raison d'être is the rejection of privileged parties with special access rights.

Another, more widely acceptable idea to combat money laundering specifically are threshold schemes.
Small payments would enjoy unlinkability while larger transactions require identification or are traceable by design~\cite{jarecki2005ProbabilisticEscrow,wust2019PRCashFast}.
The downsides of this approach include the need to agree on a threshold and, more importantly, it would require strong identities in order to prevent ``smurfing'' attacks, which split a large sum into many small payments.

We explore a different approach.
In many cases, the parties forming the anonymity set can retain some private information, which can help deanonymize other members of the set.
Collaborative deanonymization means that some parties, henceforth called \emph{witnesses}, share information on request for the purpose of solving a crime.
In a nutshell, law enforcement publicly shares information requests for specific crimes.
Then users check whether they are involved, decide whether the crime should be prosecuted, and potentially reveal private information to support deanonymization.

We argue that this approach is compatible with the peer-to-peer spirit of decentralized systems because every witness decides if she supports the investigation or not.
This limits the method to felonies that are universally disapproved, such as extortion (ransomware) or the financing of child sexual abuse.
For the method to be effective, it is not required that every witness collaborates.
Every collaborating witness reduces the search space.
Law enforcement might leverage a range of incentives to induce collaboration: alibi, altruism, bounties, and---in justified cases---force (e.\,g., seizure and use of a private key).
Unlike traffic or blockchain analyses, collaborative deanonymization does not scale, hence the risk of secret mass surveillance is small.
Moreover, as search requests are announced publicly, law enforcement can be held accountable.
The very fact that anonymity is conditional can deter crime.

In the following we develop a scenario (Sect.~\ref{sec:scenario}), formulate desiderata, and sketch protocols (Sects.~\ref{sec:backtracking_protocols} and \ref{sec:forward_tracking}) that enable collaborative deanonymization of two relevant privacy techniques: \cj{} in Bitcoin and Monero rings.
Section~\ref{sec:conclusion} concludes.

Crucially, our protocols are overlays and do not require changes to the target systems.
Similar protocols can be developed for other cryptocurrencies and privacy techniques.

\section{Scenario and Model}
\label{sec:scenario}

\begin{figure}[b]
  \centering
\begin{tikzpicture}[x=-20mm, y=0.8cm,>=stealth]

	\begin{scope}[minimum width=5mm,minimum height=5mm]
	\draw (0,0) node (A) {};

	\draw (1,1) node (B1) {};
	\draw (1,-1) node (B2) {};

	\draw (2,1.5) node (C1) {};
	\draw (2,.5) node (C2) {};
	\draw (2,-.5) node (C3) {};
	\draw (2,-1.5) node (C4) {};
	\end{scope}

	\foreach \n in {A,B1,B2,C1,C2,C3,C4}
		\draw (\n.north west)--(\n.north east) (\n.north)--(\n.south);

	\begin{scope}[inner sep=1.5pt]
		\draw [->] (A.east)--++(-.35,0) node [right, align=center, xshift=5pt] {suspicious\Aq\\cash-out\Aq};

		\draw [<-] (A)-- node [circle,fill] {} (B1.east);
		\draw [<-] (A)-- node [circle,fill] {} (B2.east);

		\draw [<-] (B1)-- node [circle,fill] {} (C1.east);
		\draw [<-] (B1)-- node [circle,fill] {} (C2.east);

		\draw [<-] (B2)-- node [circle,fill] {} (C3.east);
		\draw [<-] (B2)-- node [circle,fill] {} (C4.east);

		\foreach \n in {C1,C2,C3,C4}
		{
			\draw [->] (\n)++(.5,.2) node [circle,fill] {} -- (\n);
			\draw [->] (\n)++(.5,-.2) node [circle,fill] {} -- (\n);
		}
	\end{scope}
\end{tikzpicture}
\caption{Example entity graph of 7 ring-type transactions, $\mixin=2$.
Dots are entities, arrows denote possible payments.
Observe the exponential growth of suspects (entities on the very left).}
\label{fig:graph}
\end{figure}
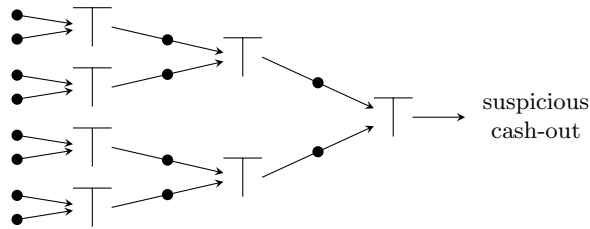

Consider a scenario where a law enforcement agency (\LEA) has identified a suspicious cash-out from a cryptocurrency address.
The objective of an investigation is to find an identifiable source, i.\,e., \emph{backtracking}.
After employing known blockchain analysis methods, like state-of-the-art clustering~\cite{goldfeder2018Whencookie}, the \LEA{} obtains an entity graph where backtracking is ambiguous only due to mixing transactions.

We model such transactions as collections of $m$ inputs and $n$ outputs.
The \LEA{} has no information about the relation.\footnote{Conversely, if the \LEA{} has some information (e.\,g.\ due to non-uniformly valued inputs and outputs), it can partition the transaction and proceed as described.}
Without loss of generality, we assume that each output of a transaction is funded by exactly one input.
Backtracking links the entity associated with the targeted $t$-th output to the entity of the funding input.
Between transactions, each input references exactly one output of a previous transaction.

We consider two of the most relevant types of mixing transactions: join-type as used in \cj{}~\cite{maxwell2013CoinJoinBitcoin} and ring-type as used in Monero~\cite{fujisaki2007TraceableRing}.%

\begin{center}
	\begin{tikzpicture}[x=3em, y=2.5ex]
		\begin{scope}[shift={(-3,0)}]
			\draw (-1,0)-- node [above] {general} (1,0);
			\draw (0,0) -- (0,-5.5);
      \small
			\draw (-0.5, -1) -- +(0, -2) node[fill=white] {$m$} -- +(0, -4);
			\draw ( 0.5, -1) -- +(0, -2) node[fill=white] {$n$} -- +(0, -4);
		\end{scope}
		\begin{scope}[]
			\draw (-1,0)-- node [above] {join-type} (1,0);
			\draw (0,0) -- (0,-5.5);
      \small
			\node at ( 0.5, -1) {$\perm(1)$};
			\node at ( 0.5, -2) {$\perm(2)$};
			\node at ( 0.5, -5) {$\perm(m)$};
			\node at (-0.5, -1) {$1$};
			\node at (-0.5, -2) {$2$};
			\node at (-0.5, -5) {$m$};
			\node at (-0.5, -3.3) {\footnotesize$\vdots$};
			\node at ( 0.5, -3.3) {\footnotesize$\vdots$};
		\end{scope}
		\begin{scope}[shift={(3,0)}]
			\draw (-1,0)-- node [above] {ring-type} (1,0);
			\draw (0,0) -- (0,-5.5);
      \small
			\node at (-0.5, -1) {$1$};
			\node at (-0.5, -2) {$2$};
			\node at (-0.5, -5) {$m$};
			\node at ( 0.5, -1) {$\select$};
			\node at (-0.5, -3.3) {\footnotesize$\vdots$};
		\end{scope}
	\end{tikzpicture}
\end{center}

Join-type transactions are formed collaboratively by $m$ parties, potentially facilitated by an intermediary such as JoinMarket~\cite{moser2016Joinme}.
We model this using $m$ inputs, each funding a distinct output ($n=m$).
A join-type transaction can then be expressed as a permutation $\perm$ on $\{1,\dots,\mixin\}$.
The \LEA's problem is to find the funding input $\perm(\target)$ of the $t$-th output.
In practice, \cj{} transactions vary in size.
A study estimates the modal value of inputs for \cj s on Bitcoin at~$m=3$~\cite{moser2016Joinme}.
Transactions with $m>10$ are rare.\footnote{A \cj{} with $m=100$ made headlines in June 2019: \url{https://www.coindesk.com/bitcoin-users-perform-what-might-be-the-largest-coinjoin-ever}.}
In contrast to join-type transactions, ring-type transactions can be formed \emph{without} the cooperation of other entities.
Moreover, a ring-type transaction does not spend all outputs referenced on its input side.
In our simplified model, ring transactions have $m$ inputs and a single output ($n=\target=1$).
The \LEA's goal is to learn the true input \select{}.\footnote{
  We depart from Monero's terminology, which calls an entire ring ``input.''
}
At the time of writing, the Monero reference implementation fixes the number of inputs to $m=11$.

For both types of mixing transactions, the anonymity of the participants is based on the observer's uncertainty about $\perm$ and $\select$, respectively.
If multiple mixing transactions are cascaded, the number of possible funding sources (\emph{suspects}) increases exponentially in the number of layers (see Figure~\ref{fig:graph}).
We propose protocols that allow the \LEA{} to reduce the number of suspects in collaboration with a subset of the involved parties.

\section{Collaborative Backtracking}
\label{sec:backtracking_protocols}

We assume an authenticated one-way communication channel from the \LEA{} to the protocol participants.
The \LEA{} uses this channel to announce inquiries on targeted transaction outputs.
Each inquiry conveys enough information so that a potential witness can decide whether she supports the request, i.\,e., whether she approves prosecution of the specific case, or not.

We further assume an unauthenticated but confidential communication channel from the witnesses to the \LEA{} and, for group testimonies, communication channels between the witnesses.
Witnesses willing to support an inquiry use these channels to give testimonies that facilitate backtracking for a single transaction.

\subsection{Individual Testimony}

An individual testimony is a protocol between a single witness and the \LEA. It results in ruling out one of the possible inputs.
Formally speaking, the witness associated with the $i$-th input should prove that $\perm(t) \neq i$ or $\select \neq i$, respectively.

For join-type transactions, the witness can testify by signing a challenge with the private keys belonging to the $i$-th input and the $j$-th output (obviously $t \neq j$).

Ring-type transactions hide the true input using traceable ring signatures~\cite{fujisaki2007TraceableRing}.
By design, these ring signatures reveal attempts to spend an input more than once.
The spending of an input yields a transaction-independent \emph{key image} that must be included in a valid signature---transactions attempting to spend the same input will contain identical key images~\cite{vansaberhagen2013CryptoNotev2}.
Let $o$ be the output of a \emph{preceding} transaction that links the witness to the suspicious transaction \tx{}.
The witness prepares a phantom transaction $\tx'$ for the \LEA{}.
It has one input referencing $o$ and one output.
The output could be invalid in order to avoid accidental inclusion in the blockchain.
For example, $\tx'$ could spend more funds than available in $o$.
Crucially, the phantom transaction unambiguously spends~$o$.
If the key image associated with $\tx'$ is different to the key image of $\tx$, it must hold that $i \neq \select$.

\subsection{Group Testimony}

The \LEA{} is interested in a single input to output relationship,
but it learns one relationship per individual testimony.
Group testimonies can avoid this unnecessary privacy loss.
Multiple witnesses controlling the set of inputs $S$ collaboratively testify $\perm(t) \not\in S$ or $\select \not\in S$, while maintaining their anonymity within~$S$.

For join-type transactions, this can be realized by signing a challenge with all $2\cdot\len{S}$ private keys belonging to the witnesses' inputs and outputs.
In the best case, all $m -1$ witnesses cooperate ($S = \set{1,\dots,m} \setminus \set{t}$) and identify the true suspect.
If $\len{S}<\mixin$ witnesses participate in the protocol, for example because private keys are deleted or witnesses unreachable, the search space is reduced to $\mixin-\len{S}$ suspects.
Join-type group testimonies retain $S$ as the anonymity set of witnesses.
Cases where $S = \set{1,\dots,m} \setminus \set{t}$ minimize the anonymity loss for witnesses when testifying that $\perm(t) \notin S$.

For ring-type transactions, it is possible to implement group testimonies with the construction of a provably spent set~\cite{wijaya2018MoneroRing,yu2019NewEmpirical}.
For example, each cooperating witness can individually form a new transaction $\tx'$ like for an individual testimony, however this time referencing not only its own input but all inputs $S$ of cooperating witnesses.
Given $\len{S}$ transactions that all have the same set of inputs $S$ and yet differing key images, the \LEA{} gains evidence that $\select \not\in S$.
If an output $o$ referenced by an input $i \in S$ is \emph{unspent} at the time of the testimony, the respective witness can achieve an anonymity set of $S$ for $o$ by referencing all~$S$ when spending $o$.
Conversely, if $o$ has already been spent in a transaction $\tx''$ with input set $S''$, the anonymity set of the witness reduces to $S \cap S''$.

Notably, each of the cooperative protocols can be executed jointly for multiple mixing transactions.
This testifies that the owners of $S$ (now generalized to the enumeration of all inputs in all transactions involved) initiated none of these transactions.
This approach is especially interesting for ring-type transactions, as larger $S$ increase the overlap with the anonymity sets of outputs that have already been spent elsewhere.

\subsection{Dealing with the Risk of False Testimonies}
\label{sub:antifake}

A general question is how much confidence the \LEA{} can place in the testimonies.
This calls for a closer look at how collaborative deanonymization can fail, and in the worst case produce false or misleading evidence.
We observe crucial differences between join-type and ring-type transactions.

Monero stores $\select$ on the blockchain, however in encrypted form. This should reduce the risk of false testimonies to the security of the cryptography used, even if private keys are leaked or stolen.

By contrast, \cj{} does not commit $\perm$ to the blockchain.
Even computationally unbounded observers cannot decide about the relation.
The resulting deniability bears a risk of false testimonies.
For example, if the perpetrator has access to the private keys of a witness, he could obtain a false alibi by signing a false input--output relation.
If the victim among the witnesses does not participate in the collaborative deanonymization, she is falsely accused.
If she does participate, the \LEA{} receives two conflicting statements.
This concentrates the suspicion on both the perpetrator and the victim, hence perpetrators have little to gain from false statements---unless their victims are unavailable.

The sketched situation highlights that parties engaging in \cj s might be exposed to physical risks under collaborative deanonymization.
A potential direction of research is to modify the protocols used for \cj{} formation in such a way that $\perm$ is committed to the blockchain at the time of the transaction.
This would obviate false accusations and reduce the incentives to attack other witnesses.
The key question to answer is under which conditions what part of~$\perm$ should be revealable.
For example, should every party commit to one relation individually?
Would a threshold scheme make sense?
Moreover, it would be desirable to make the commitment coercion-resistant.
Otherwise, the risk could reappear at the time of the \cj{} formation, rather than be mitigated.

Another approach for increasing the credibility of testimonies could be based on witnesses proving that addresses belong to the same wallet,
e.\,g., if their wallet generates addresses deterministically from a common secret.
It is an open question how such a proof can efficiently be completed without revealing more information about the wallet than necessary.

\section{Forward Tracking}%
\label{sec:forward_tracking}

A variant of the scenario presented in Section~\ref{sec:scenario} is \emph{forward tracking}.
Here, the \LEA{} has identified a suspicious origin and wishes to trace the money flow to its (current) destination or until it hits a known cash-out point.
We sketch how our approach can be adapted to this case.

\subsection{Testimonies for Forward Tracking}%
\label{sub:forward_tracking_testimonies}

\sloppy
Due to the symmetry of join-type transactions, the backtracking protocols (Sect.~\ref{sec:backtracking_protocols}) can be repurposed for forward tracking.
Since $\perm$ is bijective, testimonies which rule out assignments of $\perm$ also rule out assignments of $\perm^{-1}$.

Ring-type transactions are less straightforward.
The protocols given in Section~\ref{sec:backtracking_protocols} enable collaborating witnesses to testify that a set of inputs $S$ does not contain the funding input for a given transaction $\tx$, i.\,e., $\select \notin S$.
For the case of forward tracking, they must instead prove that only one specific suspicious input $s$ is not a funding input, i.\,e., $\select \neq s$.
Individual witnesses can accomplish this by creating a phantom transaction $\tx'$, which include all but the suspicious input $s$.
As $\tx'$ and $\tx$ share the same funding input $i$, they will produce identical key images.
By comparing the key images of $\tx$ and $\tx'$, the \LEA{} can verify that $\select \neq s$ without learning $i$.

\subsection{Blacklisting and Cover Transactions}%
\label{sub:blacklisting}

Forward tracking is related to transaction \emph{blacklisting} previously proposed (and controversially debated) as a regulatory instrument~\cite{moser2019Effectivecryptocurrency,anderson2018MakingBitcoin}.
Specifically the ``poison'' policy~\cite{moser2014RiskScoring}, where taint of a single input is propagated to all outputs, mimics the proliferation of a priori suspicion.
An interesting question is whether the threat of blacklisting can foster collaboration.
For example, the propagation policy could terminate at transactions that are whitelisted after sufficient evidence has been collected to disambiguate the entity graph (for forward and backtracking).

Forward tracking on Monero rings comes with two caveats.
First, it might be hard to decide about when to terminate (unsuccessfully), because it is often unknown whether a given output has been spent at all.
Second, the method is susceptible to cover transactions placed by a perpetrator.
Such transactions reference the investigated money flow in order to increase the search space and with it the number of witnesses needed.

Blacklisting might be a defense against this behavior because it would devalue the funds in cover transactions and thus raise the cost of creating them.
However, the effectiveness of this method as well as other defenses are open research questions.
We note that backtracking is not affected by the threat of cover transactions because funding transactions cannot be added after the spending transaction.

\section{Conclusion and Outlook}%
\label{sec:conclusion}

We have outlined a novel way to investigate criminal money flows in cryptocurrencies even if the perpetrators use anonymization techniques.
Our approach requires collaboration of witnesses, which keeps the method costly enough to prevent mass surveillance or the prosecution of petty crimes.
Specifically, we have given protocols for backtracking and forward tracking of \cj{} transactions in Bitcoin as well as Monero rings.
Several techniques ensure that the information shared with law enforcement can be limited to the necessary minimum.
The new risk of false accusations has been discussed.
A general consequence of collaborative deanonymization is that old private keys remain sensitive even if they do not control any funds anymore.

We shall also pinpoint future work.
Obviously, the protocols for secure testimonies need to be further developed and their properties formalized and proven.
A proof-of-concept implementation for the most relevant types of mixing transactions could demonstrate the practicality of our approach.
Whether and under which condition \LEA s can deploy collaborative deanonymization, must be subject of more interdisciplinary work with legal scholars.
Adapting the approach to less common types of mixing transactions (see for instance Table~1 of \cite{heilman2017TumbleBitUntrusted} for an overview) would help to complete the picture.

The topic also lends itself to economic studies.
One could investigate the incentives of witnesses to collaborate, presumably with cooperative game theory~\cite{arce2018PricingAnonymity}.
In addition, potential knock-on effects on the participation in mixing transactions call for a model in the tradition of competitive game theory~\cite{abramova2017MixingCoins}.

Two broader technical directions are to explore collaborative deanonymization for anonymous communication systems, and to research deniable privacy techniques, which could protect potential witnesses from any pressure to testify or release deanonymizing information.

In summary, collaborative deanonymization appears not only under-resear\-ched, but also under-estimated for its potential to balance the conflicting goals of privacy and law enforcement in future digital currency systems. This short paper sets out to make a case for this promising tool.

\small
\section*{Acknowledgements}

We thank our colleagues
Michael Fröwis,
Malte Möser,
Tim Ruffing,
and a number of anonymous reviewers for helpful discussions of earlier versions of this work.
Rainer Böhme's and Patrik Keller's work on this topic is supported by the Austrian FFG’s KIRAS programme under project VIRTCRIME.

\bibliographystyle{splncs04}
\bibliography{main}

\end{document}